\documentclass[12pt]{article}
\usepackage{epsfig,graphicx}

\begin{document}

\begin{center}

\vspace*{1.0cm}

{\Large \bf{Search for $2\beta$ decays of $^{96}$Ru and $^{104}$Ru by ultra-low background
HPGe $\gamma$ spectrometry at LNGS: final results}}

\vskip 1.0cm

{\bf
P.~Belli$^{a}$,
R.~Bernabei$^{a,b,}$\footnote{Corresponding author.
   {\it E-mail address:} rita.bernabei@roma2.infn.it (R.~Bernabei).},
F.~Cappella$^{c,d}$,
R.~Cerulli$^{e}$,
F.A.~Danevich$^{f}$,
S.~d'Angelo$^{a,b}$,
A.~Incicchitti$^{c,d}$,
G.P.~Kovtun$^{g}$,
N.G.~Kovtun$^{g}$,
M.~Laubenstein$^{e}$,
D.V.~Poda$^{f}$,
O.G.~Polischuk$^{c,f}$,
A.P.~Shcherban$^{g}$,
D.A.~Solopikhin$^{g}$,
J.~Suhonen$^{h}$,
V.I.~Tretyak$^{f}$
}

\vskip 0.3cm

$^{a}${\it INFN, sezione Roma ``Tor Vergata'', I-00133 Rome, Italy}

$^{b}${\it Dipartimento di Fisica, Universit\`a di Roma ``Tor Vergata'', I-00133 Rome, Italy}

$^{c}${\it INFN, sezione Roma ``La Sapienza'', I-00185 Rome, Italy}

$^{d}${\it Dipartimento di Fisica, Universit\`a di Roma ``La Sapienza'', I-00185 Rome, Italy}

$^{e}${\it INFN, Laboratori Nazionali del Gran Sasso, I-67100 Assergi (AQ), Italy}

$^{f}${\it Institute for Nuclear Research, MSP 03680 Kyiv, Ukraine}

$^{g}${\it National Science Center ``Kharkiv Institute of Physics and Technology'', 61108 Kharkiv, Ukraine}

$^{h}${\it Department of Physics, University of Jyv\"{a}skyl\"{a}, P.O. Box 35 (YFL), FI-40014 Finland}

\end{center}

\vskip 0.5cm

\begin{abstract}

An experiment to search for double $\beta$ decay processes in $^{96}$Ru and $^{104}$Ru,
which are accompanied by $\gamma$ rays, has been realized in the underground Gran Sasso
National Laboratories of the I.N.F.N. (Italy). Ruthenium samples with masses of
$\approx(0.5-0.7)$~kg were measured with the help of ultra-low background high purity Ge
$\gamma$ ray spectrometry. After 2162 h of data taking the samples were deeply purified to reduce the
internal contamination of $^{40}$K. The last part of the data has been 
accumulated over 5479 h.
New improved half life limits on $2\beta^+/\varepsilon\beta^+/2\varepsilon$ processes in $^{96}$Ru
have been established on the level of $10^{20}$~yr, in particular for decays to the ground state of
$^{96}$Mo:
$T_{1/2}^{2\nu 2\beta^+}\geq1.4\times 10^{20}$~yr,
$T_{1/2}^{2\nu \varepsilon\beta^+}\geq8.0\times10^{19}$~yr and
$T_{1/2}^{0\nu 2K}\geq1.0\times 10^{21}$~yr
(all limits are at 90\% C.L.).
The resonant neutrinoless double electron captures to the 2700.2 keV and 2712.7 keV excited states
of $^{96}$Mo are restricted as:
$T_{1/2}^{0\nu KL}\geq2.0\times10^{20}$~yr and
$T_{1/2}^{0\nu 2L}\geq3.6\times 10^{20}$~yr, respectively.
Various two neutrino and neutrinoless $2\beta$ half lives of $^{96}$Ru
have been estimated in the framework of the QRPA approach.
In addition, the $T_{1/2}$ limit for $0\nu2\beta^-$ transitions of 
$^{104}$Ru to
the first excited state of $^{104}$Pd has been set as $\geq6.5\times10^{20}$~yr.

\end{abstract}

\vskip 0.4cm


\vskip 0.4cm


\section{Introduction}

Double beta ($2\beta$) decay is a process of transformation of a nucleus 
$(A,Z)$ either to $(A,Z+2)$
with simultaneous emission of two electrons ($2\beta^-$ decay) or to $(A,Z-2)$ through one of the 
following ways:
emission of two positrons ($2\beta^+$),
capture of electron and emission of positron ($\varepsilon\beta^+$) or
double electron capture ($2\varepsilon$).
The two neutrino ($2\nu$) double $\beta$ decay, %transfromation $(A,Z)$ $\to$ $(A,Z+2)$
in which two (anti)neutrinos are also emitted, is allowed in the Standard Model (SM);
however, being a second order process in the weak interactions, it is characterized
by very long half lives in the range of $10^{18} - 10^{24}$ yr \cite{2bRev}.
There are 35 known nuclei candidates for $2\beta^-$ and 34 candidates for
$2\beta^+/\varepsilon\beta^+/2\varepsilon$ decays \cite{2bTab}.
To-date, two neutrino $2\beta$ decays are observed for several $2\beta^-$ decaying nuclei 
(see reviews \cite{2bTab,Bar10} and recent original works \cite{Ack11,Gan12}),
while indications on double electron capture have been obtained for $^{130}$Ba
in geochemical experiments \cite{Mes01,Puj09}.

The neutrinoless ($0\nu$) mode of the $2\beta$ decay is forbidden in the SM because it violates the
lepton number by 2 units. It is, however, naturally expected in many SM extensions which
describe the neutrino as a Majorana particle with non-zero mass. The neutrino oscillation
experiments indicate that the neutrinos are massive. Nevertheless, since they are sensitive 
to the
difference in $\nu$ masses, the absolute $\nu$ mass scale is unknown \cite{Dor08}.
The $0\nu2\beta$ decay is considered a powerful tool to check lepton number conservation,
to determine the absolute $\nu$ masses and their hierarchy, to establish the nature of the neutrino
(Majorana or Dirac particle), to find a possible contribution of right-handed admixtures to
weak interaction and the existence of Nambu-Goldstone bosons (Majorons).
A particular analysis of data on $^{76}$Ge provided an evidence for $0\nu2\beta$ decay 
\cite{Kla06}; several experiments with the aim to test it and to explore the inverted 
hierarchy of the Majorana neutrino mass region ($m_{\nu}\sim 0.1-0.05$ eV) are now in
progress or under development \cite{2bRev}.
Studies of neutrinoless $2\beta^-$ and $2\beta^+/\varepsilon\beta^+/2\varepsilon$ decays are mutually
complementary, helping to distinguish contributions from the neutrino mass and right-handed admixture
mechanisms \cite{Hir94}.

$^{96}$Ru is one of the only six isotopes where the decay with emission of two positrons is
allowed \cite{2bTab} thanks to the high energy release: $Q_{2\beta} = (2714.51\pm0.13)$ keV \cite{Eli11}.
It has also a quite big natural abundance: $\delta = 5.54$\% \cite{Ber11}.
Moreover, in case of capture of two electrons from the $K$ and $L$ shells
(the binding energies are $E_K = 20.0$ keV, $E_{L1} = 2.9$ keV \cite{ToI98})
or both from the $L$ shell, the decay energies $(2691.61\pm0.13)$ keV
and $(2708.71\pm0.13)$ keV are close to the energy of the excited levels of $^{96}$Mo 
($E_{exc} = 2700.21$ and 2712.68 keV \cite{Abr08}). 
Such a situation
could give rise to a resonant enhancement of the neutrinoless $KL$ and $2L$
capture to the corresponding level of the daughter nucleus as a result of the energy degeneracy
\cite{r2e0n}\footnote%
{In accordance with older atomic masses \cite{Aud03}, the energy release 
$Q_{2\beta} = (2718\pm8)$ keV gave the decay energies for $KL$ and $2L$ captures as
$(2695\pm8)$ keV and $(2712\pm8)$ keV, respectively, compatible within uncertainties to the
energies of the $^{96}$Mo excited levels (2700.2 and 2712.7 keV), and $^{96}$Ru was considered as a 
very promising candidate in looking for resonant $0\nu2\varepsilon$ captures.
After the recent high-precision measurement of Ref. \cite{Eli11}: %$Q_{2\beta} = (2714.51\pm0.13)$ keV, 
$^{96}$Ru is no more considered a 
promising candidate to search for this process.}.

In addition, another isotope of ruthenium, $^{104}$Ru, is potentially
unstable with respect to the $2\beta^-$ decay ($Q_{2\beta} = (1301.2\pm2.7)$ keV 
\cite{Wan12}, $\delta = 18.62$\%).
The decay schemes of $^{96}$Ru and $^{104}$Ru are shown in Fig.~1.

\begin{figure}[htb]
\begin{center}
\resizebox{0.54\textwidth}{!}{\includegraphics{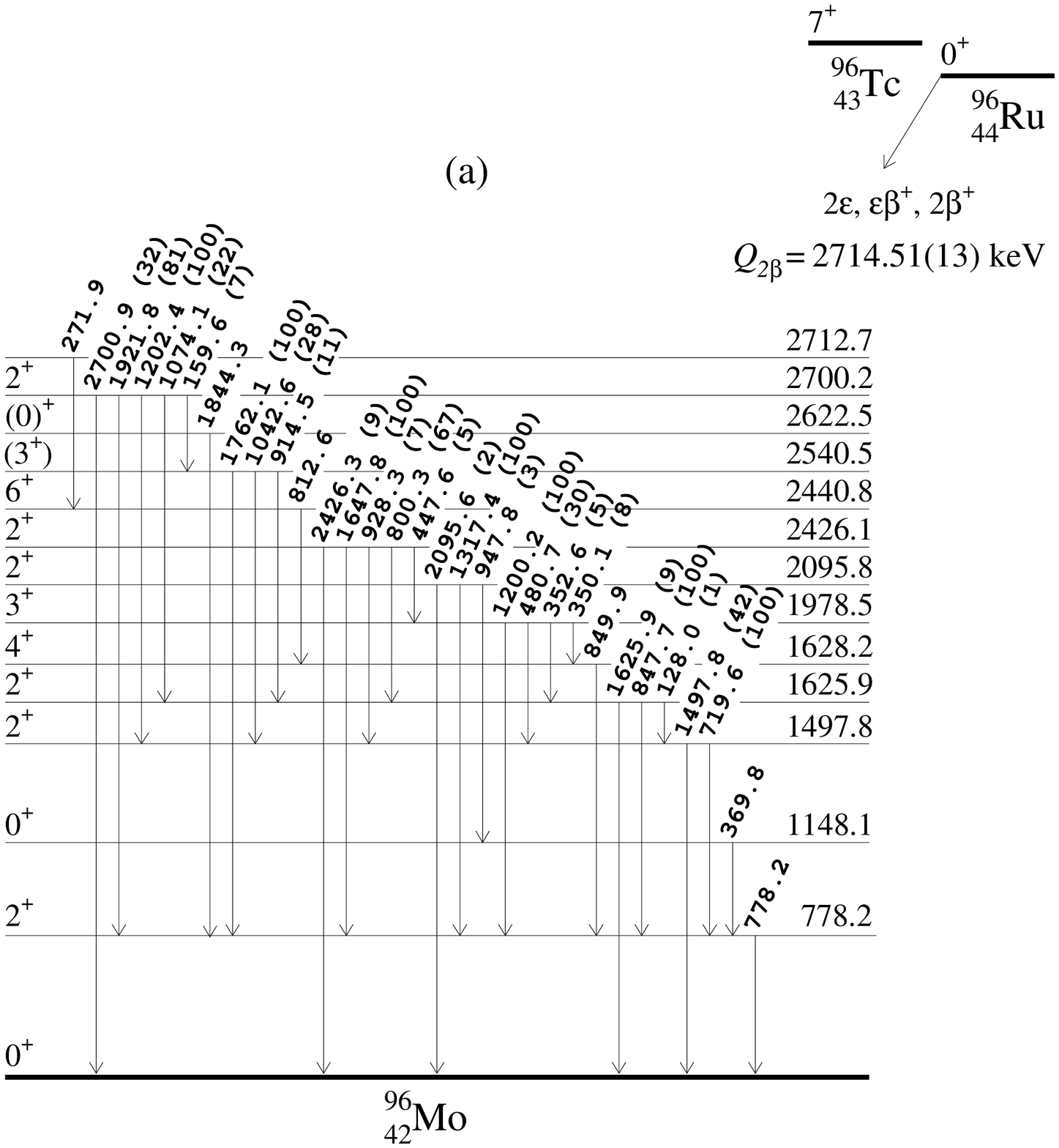}}~~~
\resizebox{0.36\textwidth}{!}{\includegraphics{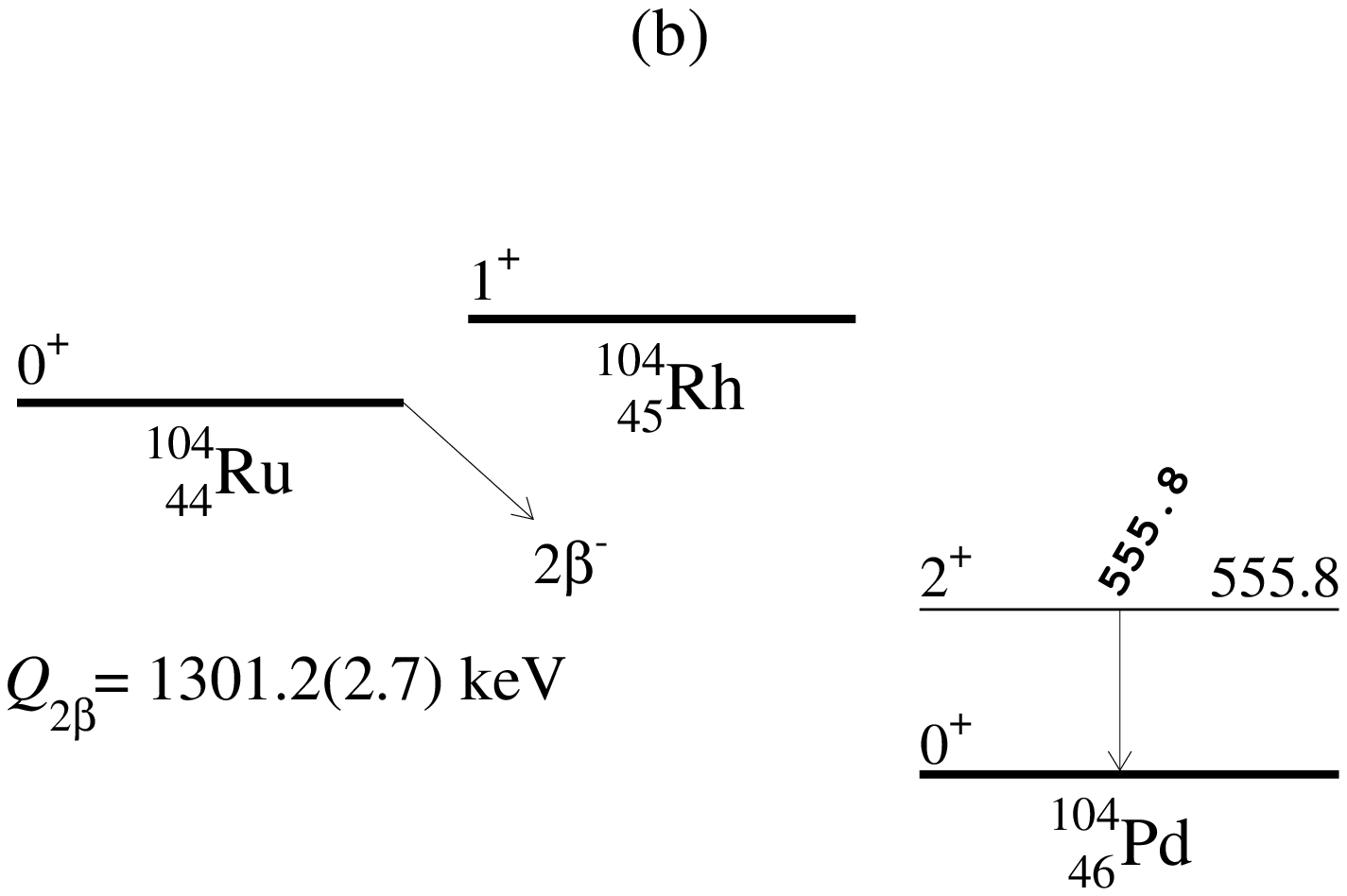}}
\end{center}
\caption{Decay schemes of $^{96}$Ru (a) and $^{104}$Ru (b).
Energies of the excited levels and emitted $\gamma$ quanta
are in keV. The relative intensities of $\gamma$ quanta are given in parentheses
\cite{ToI98,Abr08,Bla07}.}
\end{figure}

Despite the high energy release and the high abundance, only one search for 
$2\beta^+/\varepsilon\beta^+$ processes
in $^{96}$Ru was performed in 1985, giving $T_{1/2}$ limits on the level of
$10^{16}$ yr \cite{Nor85}. The efforts were renewed only in 2009, when a Ru sample with a
mass of 473 g was measured for 158 h with an HPGe detector (468 cm$^3$) in the underground 
conditions of the
Gran Sasso National Laboratories (Laboratori Nazionali del Gran Sasso, LNGS)
of the I.N.F.N. (3600 m w.e.) \cite{Bel09} (an updated statistics of 2162
h was then reported in \cite{Bel10}). The achieved sensitivity for the
$2\beta^+/\varepsilon\beta^+/2\varepsilon$ decays was $10^{18}-10^{19}$ yr;
for several modes of $2\beta$ decay of $^{96}$Ru (and $^{104}$Ru) $T_{1/2}$ limits were established
for the first time.
A search for $2\beta$ decays of Ru was also performed in the HADES
underground laboratory (500 m w.e.) where a sample of Ru with mass of 149 g was measured
during 2592 h; $T_{1/2}$ limits were obtained on the level of $10^{19}$ yr \cite{And12}.

Our previous measurements \cite{Bel09,Bel10} showed that the used Ru sample was contaminated
by $^{40}$K at $\simeq3$ Bq/kg, and better results are possible only with purified Ru.
Here we report the final results of the 
search for $2\beta^+/\varepsilon\beta^+/2\varepsilon$ processes in $^{96}$Ru and for
$2\beta^-$ decay in $^{104}$Ru obtained with a purified sample of Ru (720 g)
in measurements during 5479 h.

\section{Purification of ruthenium and low background measurements}

The ruthenium (with natural isotopic composition) of 99.99\% grade
produced by powder metallurgy was provided by Heraeus \cite{Heareus}.
At the first stage \cite{Bel09,Bel10}, the Ru sample with total mass of 473 g was in form of pellets
(50 tablets $\oslash16\times5$~mm, density $\approx8.7$ g/cm$^3$).
The analysis of the data showed a high level of $^{40}$K contamination in the ruthenium (3.4~Bq/kg),
and for further measurements the Ru sample with an increased total mass of 946 g was purified by
an electron beam melting method.

The ruthenium was divided into five parts, each with a mass of almost 0.2 kg, which were slowly melted
(to avoid intensive sprinkling) using an electron beam and kept in a liquid state under vacuum (0.01 -- 
0.05 Pa)
for $\simeq 1.5 - 2$ h.
The purification occurs through the evaporation of the impurities from the melted ruthenium.
More details about the purification process can be found in \cite{Bob11}.
As a result, five Ru samples in form of oval disks (totally 719.5 g, density $\approx10.0$ g/cm$^3$)
were obtained.

The purified ruthenium samples were measured over
5479 h in the GeMulti set-up (made of four HPGe detectors; $\simeq225$ cm$^3$ each one) installed deep
underground at the LNGS.
The detectors are surrounded by a passive shield made of low
radioactivity copper ($\simeq5$ cm thick) and low radioactivity lead ($\simeq25$ cm).
The set-up was continuously flushed with high purity nitrogen to remove radon.
The typical energy resolution of the detectors is 2.0 keV at the 1332.5 keV line of $^{60}$Co. The energy
spectra without samples were accumulated with the GeMulti spectrometer over 7862 h. The
results of the measurements are presented in Fig.~2, where the effect of the purification is clearly
visible (here and in the following, the spectra of the GeMulti set-up are the sum of the 
spectra of the 4 individual HPGe detectors). 
Table~1 gives a summary of the measured
radioactive contaminations in the used Ru before and after the purification process.
The purification allowed us to decrease the $^{40}$K contamination of $\simeq20$ times;
the contamination by $^{226}$Ra was also suppressed by $\simeq6$ times, while 
the activity of $^{106}$Ru was decreased by $\simeq5$ times due to decay with the 
half life $T_{1/2}=371.8$ d \cite{De2008}.

\begin{figure}[htb]
\begin{center}
\resizebox{0.54\textwidth}{!}{\includegraphics{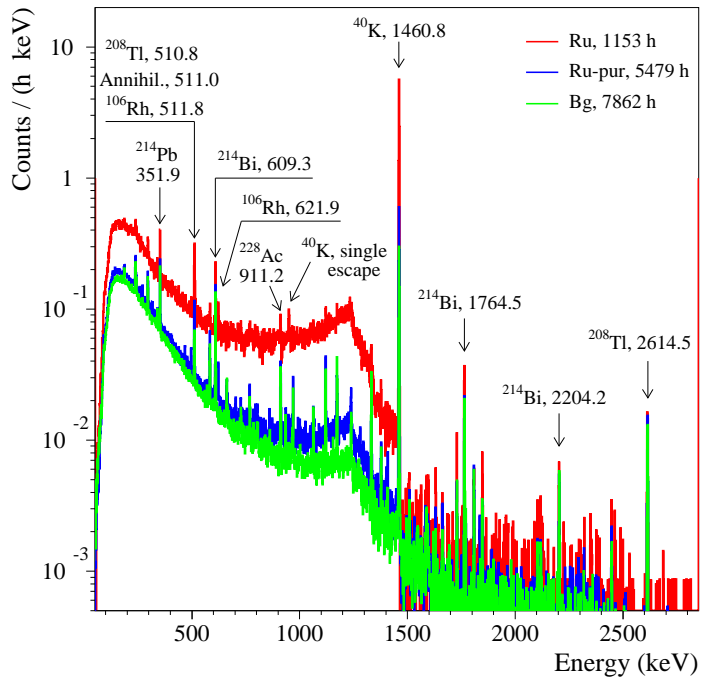}}
\end{center}
\caption{(Color on-line) The energy spectra above 20 keV accumulated with the initial Ru sample
over 1153 h (Ru) and with the purified Ru over 5479 h (Ru-pur) in comparison with the background (Bg)
of the GeMulti ultra-low background HPGe $\gamma$ spectrometer measured over 7862 h.
The energies of $\gamma$ lines are in keV.}
\end{figure}

\begin{table}[htb]
\caption{Radioactive contamination of the Ru sample used in \cite{Bel09} (473 g, 158 h) and of the purified Ru
sample (720 g, 5479 h, measured here). For comparison, the results of the sample used in \cite{And12} (149 g,
2592 h) are also presented. The limits are given at 90\% C.L. (95\% C.L. for \cite{And12}).
Activity of $^{103}$Ru ($T_{1/2}=39.26$ d \cite{ToI98}) is quoted for the beginning of the
present measurements.}
\begin{center}
\begin{tabular}{lllll}
\hline
Chain      & Nuclide        & \multicolumn{3}{c}{Activity (mBq/kg)} \\
\cline{3-5}
~          & ~              & Ru \cite{Bel09} & Purified Ru & Ru \cite{And12} \\
\hline
$^{228}$Th & $^{228}$Ra     & $\leq7.1$       & $\leq1.0$   & $8.7\pm0.7$ \\
~          & $^{228}$Th     & $\leq3.4$       & $1.4\pm0.4$ & $8.8\pm0.6$ \\
~ & ~ & ~ & ~ & ~ \\
$^{235}$U  & $^{235}$U      & $\leq6.9$       & $\leq4.0$   & -- \\
~ & ~ & ~ & ~ & ~ \\
$^{238}$U  & $^{234}$Th     & $\leq390$       & --          & $\leq36$ \\
~          & $^{234}$Pa$^m$ & $\leq260$       & $\leq23$    & -- \\
~          & $^{226}$Ra     & $6.4\pm1.7$     & $1.0\pm0.3$ & $14.6\pm0.7$ \\
~          & $^{210}$Pb     & --              & --          & $\leq100$ \\
~ & ~ & ~ & ~ & ~ \\
~          & $^{40}$K       & $3400\pm600$    & $153\pm4$   & $169\pm7$ \\
~          & $^{60}$Co      & $\leq1.7$       & $\leq0.1$   & $\leq0.2$ \\
~          & $^{137}$Cs     & $\leq2.6$       & $\leq0.1$   & $\leq0.2$ \\
~ & ~ & ~ & ~ & ~ \\
~          & $^{103}$Ru     & --              & $3.3\pm0.7$ & -- \\
~          & $^{106}$Ru     & $24\pm7$        & $5.0\pm0.6$ & $\leq1.7$ \\
\hline
\end{tabular}
\end{center}
\end{table}

\section{New experimental $T_{1/2}$ limits on $2\beta$ decay of ruthenium}

We did not observe any peak in the spectra accumulated with the ruthenium sample which
could be unambiguously attributed to the $2\beta$ processes in $^{96}$Ru and $^{104}$Ru.
Therefore only lower half life limits are given using the formula:

\begin{equation}
\lim T_{1/2} = N \cdot \eta \cdot t \cdot \ln 2 / \lim S,
\end{equation}

\noindent where $N$ is the number of potentially 2$\beta$ unstable nuclei in the Ru sample,
$\eta$ is the detection efficiency, $t$ is the measuring time, and
$\lim S$ is the number of events of the effect searched for which can be excluded at a
given confidence level (C.L.; all the limits in the present study are given at 90\% C.L.).
The efficiency of the detectors for the double $\beta$ processes in $^{96}$Ru and $^{104}$Ru
has been calculated by using the EGS4 code \cite{EGS4} with initial kinematics given by the
DECAY0 event generator \cite{Decay0}.
The procedure of the analysis, in particular in determining the $\lim S$ values,
is well described in \cite{Bel09}.

\subsection{Search for $2\beta^+$ decay of $^{96}$Ru}

Only the ground state of $^{96}$Mo can be populated in the $2\beta^+$ decay of $^{96}$Ru, and thus only
annihilation $\gamma$ quanta with energy 511.0 keV could be registered by our detectors.
A possible extra rate in the annihilation peak in the spectrum accumulated with the purified Ru
sample (see Fig.~3) could be related to the $2\beta^+$ (and $\varepsilon\beta^+$) 
decay of $^{96}$Ru.

\begin{figure}[htb]
\begin{center}
\resizebox{0.54\textwidth}{!}{\includegraphics{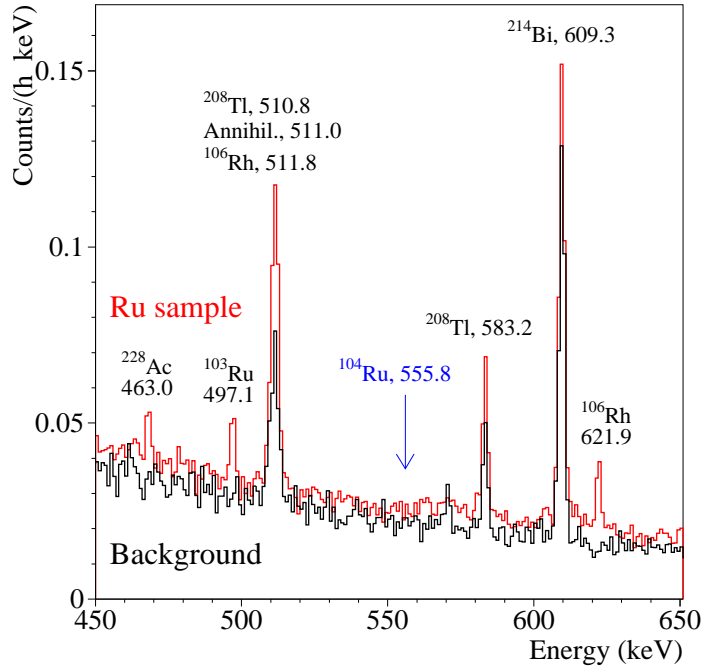}}
\end{center}
\caption{(Color on-line)  Fragment of the energy spectra accumulated
with the ruthenium sample over 5479 h (Ru sample) and without the sample
over 3362 h (Background; see footnote $^3$) in the vicinity of the 
annihilation peak.}
\end{figure}

The area of the annihilation peak in the measurements with the purified Ru during 5479 h
is equal to $(1461\pm39)$ counts, while in the background spectrum it is $(535\pm27)$ counts
during 3362 h\footnote{We use here the last series of the background measurements with the
GeMulti set-up close to the data accumulated with the purified Ru sample.};
this gives $(589\pm58)$ counts of the extra events.
The excess is explained by the following contributions:

1) the 511.8 keV $\gamma$ line from $^{106}$Rh which is the daughter radionuclide of the cosmogenic
$^{106}$Ru; this contribution is estimated to be $(433\pm52)$ counts using the supplementary 
$^{106}$Rh peak
at 621.9 keV with an area $(197\pm24)$ counts, taking into account the different yields of these $\gamma$
quanta per decay ($\gamma_{512}=20.4\%$ and $\gamma_{622}=9.93\%$ \cite{ToI98})
and their detection efficiencies ($\eta_{512}=3.0\%$ and $\eta_{622}=2.8\%$);

2) the 510.8 keV $\gamma$ line from the $^{208}$Tl contamination in the Ru sample;
this contribution is estimated in a similar way as before to be $(49\pm11)$ counts 
using the area
of the 583.2 keV peak of $^{208}$Tl
(with subtraction of the contribution from the corresponding background peak);

3) the $e^+e^-$ pairs created by the 1460.8 keV $\gamma$ quanta emitted in the $^{40}$K decay;
this contribution is estimated to be $(165\pm4)$ counts using the area of the 1460.8 keV peak,
$(5266\pm132)$ counts (after subtraction of the corresponding background peak),
and results of simulation with the EGS4, which give a ratio of 1:31.9
between the 511.0 keV and
1460.8 keV peaks from $^{40}$K in our measurements.

The difference between the measured number of events in the 511.0 keV peak and the
estimated contributions from the known sources, $(-58\pm79$) counts, could eventually be ascribed to 
the effect
searched for. Obviously, there is no evidence of $2\beta^+$ (and $\varepsilon\beta^+$) decay of
$^{96}$Ru to the ground state of $^{96}$Mo. In accordance with the Feldman-Cousins procedure
\cite{Fel98}, it results in a limit $\lim S = 79$ counts for the effect, which can be
excluded at 90\% C.L.
Taking into account the calculated efficiency for $2\beta^+$ processes
(10.36\% for $2\nu$ mode and 10.31\% for $0\nu$) and the number of the $^{96}$Ru nuclei
($N = 2.38\times10^{23}$), this gives:
\begin{equation}
T_{1/2}^{2\nu 2\beta^+}({\rm g.s.} \rightarrow {\rm g.s.}) \geq1.4\times 10^{20}~{\rm yr}, 
\end{equation}
\begin{center}
$T_{1/2}^{0\nu 2\beta^+}$(g.s. $\rightarrow$ g.s.) $\geq1.3\times 10^{20}$~yr.
\end{center}

In case if the observed number of events is less than the expected background and thus
the estimated effect is negative, G.J.~Feldman and R.D.~Cousins recommended \cite{Fel98} 
to give, in addition to the upper limit, also the so-called sensitivity of the experiment
defined as ``the average upper limit that would be obtained by an ensemble of experiments 
with the expected background and no true signal''. Using the total number of events in the 
range of $509-514$ keV as the background ($B=2522$ counts) and extrapolating Table XII of 
\cite{Fel98} (in terms of $\sqrt{B}$), we obtain $\lim S_s = 93$ counts at 90\% C.L.
This gives the ``sensitivity'' $T_{1/2s}$ value e.g. for the $0\nu 2\beta^+$ decay as
$T_{1/2s}^{0\nu 2\beta^+}$(g.s. $\rightarrow$ g.s.) $\geq1.1\times 10^{20}$~yr,
which is very close to the obtained above value of $1.3\times 10^{20}$~yr. 
We accept the values (2) as the final ones (also to compare with results from other
experiments where only the upper limits are given).

In addition to the analysis of the usual 1-dimensional spectrum, the GeMulti set-up with its 4
HPGe detectors has the possibility to use coincidences between different detectors for 
$\gamma$ quanta
emitted simultaneously (annihilation $\gamma$ quanta in $2\beta^+$ and $\varepsilon\beta^+$ decays,
and $\gamma$'s from cascades in the deexcitation of the excited $^{96}$Mo levels).
The set-up, with and without Ru sample, has been operated 
in coincidence mode over
5479 h and 2490 h, respectively. The procedure of the analysis is the same as described recently
in \cite{Bel10a}, where a 2-dimensional spectrum was used to detect the $2\nu 2\beta^-$ transition
of $^{100}$Mo to the first excited $0^+_1$ state of $^{100}$Ru.
So, fixing the energy of one of the detectors to the expected $511\pm3$ keV
(in accordance with the energy resolution for the annihilation peak), we observe the
coincidence peak at the corresponding energy $511\pm3$ keV.
There are 18(4) counts in both the spectra. The Monte Carlo simulations give the efficiencies
to get coincidences of two $\gamma$ quanta with energy 511.0 keV:
0.30\% for $0\nu2\beta^+$ decay of $^{96}$Ru and $8\times10^{-5}$\% for $^{40}$K (which gives the
biggest contribution of 2 counts during 5479 h).
The difference between the observed and the expected number of counts $(-24\pm10)$
corresponds to $\lim S$ = 3.3 counts at 90\% C.L. that results in
$T_{1/2}^{0\nu 2\beta^+}$(g.s. $\rightarrow$ g.s.) $\geq9.3\times 10^{19}$~yr.
This value is comparable but slightly lower than that obtained above from the analysis of the
1-dimensional spectrum. This concerns also other limits obtained from the analysis
of coincidences: while they are comparable with those derived from the 1-dimensional spectrum,
in general they are lower due to a lower detection efficiency.

\subsection{$\varepsilon\beta^+$ processes in $^{96}$Ru}

The limit obtained above for the $2\beta^+$ decay
considering the 511.0 keV peak ($\lim S = 79$ counts at 90\% C.L.) can be used to estimate a half
life limit on the $\varepsilon\beta^+$ decay to the ground state of $^{96}$Mo.
Taking into account the efficiencies for the $\varepsilon\beta^+$ decay of $^{96}$Ru
(6.12\% for $2\nu$ mode and 5.89\% for $0\nu$), we obtain:
\begin{center}
$T_{1/2}^{2\nu \varepsilon\beta^+}$(g.s. $\rightarrow$ g.s.) $\geq8.0\times 10^{19}$~yr,\\
$T_{1/2}^{0\nu \varepsilon\beta^+}$(g.s. $\rightarrow$ g.s.) $\geq7.7\times 10^{19}$~yr.
\end{center}

In addition to the transition to the ground state, also few excited levels of $^{96}$Mo
can be populated in $\varepsilon\beta^+$ decay of $^{96}$Ru (up to the level $2^+$, 1625.9 keV).
To estimate the number of events ($\lim S$), the experimental energy spectrum was fitted in different
energy intervals with the sum of components representing the background (internal $^{40}$K,
U/Th, external $\gamma$ from the details of the set-up) and the EGS4-simulated models for $2\beta$
processes in $^{96}$Ru.
The used fitting approach is described in detail in Ref. \cite{Bel09}.

For example, in the case of the transition to the 778.2 keV level of $^{96}$Mo, a peak at 778.2 keV
should be present in the energy spectrum accumulated with the Ru sample.
To estimate an upper limit the spectrum was fitted in the
energy interval $(744-799)$ keV using a model made of four Gaussian functions at
the energies of 768.4 keV ($\gamma$ peak from $^{214}$Bi), 786.0 keV ($^{214}$Pb), 794.9 keV ($^{228}$Ac)
and 778.2 keV (the expected effect) with the energy resolution FWHM = 2.0 keV, and a linear function
representing the background (see Fig.~4). The fit using the chi-square method ($\chi^2$/n.d.f. =
38.8/42 = 0.92, where n.d.f. is number of degrees of freedom) results in a peak area of
 $S = (-16.9\pm14.6)$ counts, which gives no evidence for the
effect. In accordance with the procedure \cite{Fel98}, one should take 10.5 counts which
can be excluded at 90\% C.L. (fits in other energy intervals give close results).
Taking into account the detection efficiency
(2.38\%), we have obtained the following limit:
\begin{center}
$T^{(2\nu + 0\nu) \varepsilon\beta^+}_{1/2}$ (g.s. $\rightarrow$ 2$^+$, 778.2 keV) $\geq 2.3 \times 10^{20}$ yr.
\end{center}

\begin{figure}[htb]
\begin{center}
\resizebox{0.54\textwidth}{!}{\includegraphics{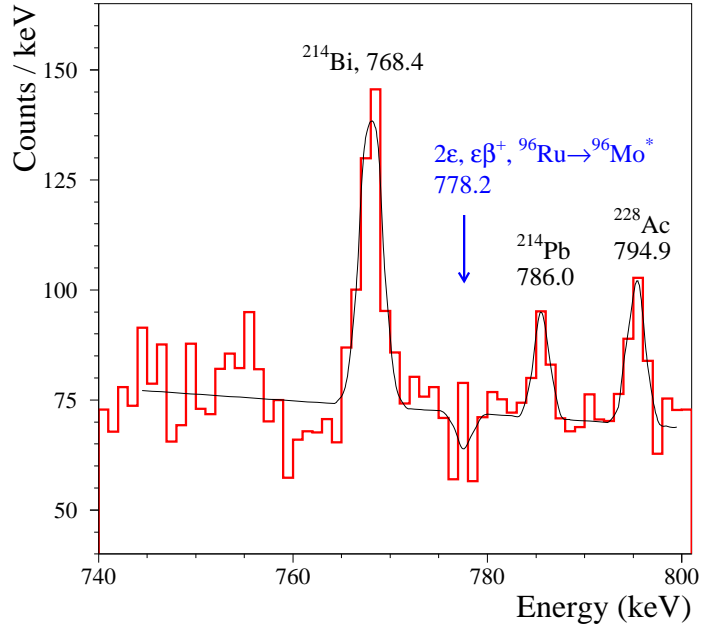}}
\end{center}
\caption{(Color on-line) Fragment of the energy spectrum accumulated with the purified ruthenium
sample over 5479 h with the ultra-low background HPGe $\gamma$ spectrometer. The fit is shown by 
solid
line. The arrow shows the energy of the peak expected in decay of $^{96}$Ru through the $2^+, 778.2$ keV level
of $^{96}$Mo. The energies of $\gamma$ lines are in keV.}
\end{figure}

Similar fits allow us to set limits on possible transitions to other excited levels
in the $\varepsilon\beta^+$ decay of $^{96}$Ru; obtained results are listed in Table~2.

\begin{table*}[htbp]
\caption{The half life limits on 2$\beta$ processes in $^{96}$Ru and $^{104}$Ru
isotopes together with theoretical predictions. The energies of the $\gamma$
lines, which were used to set the $T_{1/2}$ limits, are listed
in column 4 with the corresponding detection efficiencies ($\eta$) in column 5.
The theoretical $T_{1/2}$ values for $0\nu$ mode are given for $m_{\nu}$ = 1 eV.}
\begin{center}
\resizebox{0.96\textwidth}{!}{
\begin{tabular}{llllllll}
\hline
~                    & ~             & Level of       & ~              & ~      & \multicolumn{2}{l}{Experimental limits,}        & ~ \\
\multicolumn{2}{l}{Process of decay} & daughter       & $E_{\gamma}$   & $\eta$ & \multicolumn{2}{l}{$T_{1/2}$ (yr) at 90\% C.L.} & Theoretical estimations, \\
\cline{6-7}
~                    & ~             & nucleus        & (keV)          & (\%)   & Present                & Ref. \cite{And12}      & $T_{1/2}$ (yr)           \\
~                    & ~             & (keV)          & ~              & ~      & work                   & ~                      & ~                        \\
\hline
~                    & ~             & ~              & ~              & ~      & ~                      & ~                      & ~                        \\
\multicolumn{8}{l}{$^{96}$Ru $\to$ $^{96}$Mo} \\
2$\beta^+$           & 2$\nu$        & g.s.           & 511.0          & 10.36  & $\geq1.4\times10^{20}$ & $\geq5.0\times10^{19}$ & $1.2\times10^{26}-1.0\times10^{27}$ \\
~                    & 0$\nu$        & g.s.           & 511.0          & 10.31  & $\geq1.3\times10^{20}$ & $\geq5.0\times10^{19}$ & $5.9\times10^{27}-1.0\times10^{28}$ \\
$\varepsilon\beta^+$ & 2$\nu$        & g.s.           & 511.0          & 6.12   & $\geq8.0\times10^{19}$ & $\geq5.5\times10^{19}$ & $2.0\times10^{21}-2.3\times10^{22}$ \\
~                    & ~             & $2^+$ 778.2    & 778.2          & 2.38   & $\geq2.3\times10^{20}$ & $\geq2.7\times10^{19}$ & $1.3\times10^{27}-1.2\times10^{31}$ \\
~                    & ~             & $0^+$ 1148.1   & 778.2          & 2.26   & $\geq2.1\times10^{20}$ & $\geq1.8\times10^{19}$ & $6.1\times10^{24}-1.9\times10^{26}$ \\
~                    & ~             & $2^+$ 1497.8   & 778.2          & 1.56   & $\geq1.5\times10^{20}$ & $\geq1.3\times10^{19}$ & $2.1\times10^{33}-1.6\times10^{37}$ \\
~                    & ~             & $2^+$ 1625.9   & 847.7          & 1.96   & $\geq3.1\times10^{20}$ & $\geq1.6\times10^{19}$ & $>3.4\times10^{38}$ \\
~                    & 0$\nu$        & g.s.           & 511.0          & 5.89   & $\geq7.7\times10^{19}$ & $\geq5.5\times10^{19}$ & $5.0\times10^{26}-1.0\times10^{27}$ \\
~                    & ~             & $2^+$ 778.2    & 778.2          & 2.39   & $\geq2.3\times10^{20}$ & $\geq2.6\times10^{19}$ & -- \\
~                    & ~             & $0^+$ 1148.1   & 778.2          & 2.26   & $\geq2.1\times10^{20}$ & $\geq1.8\times10^{19}$ & $(1.0-8.2)\times10^{28}$ \\
~                    & ~             & $2^+$ 1497.8   & 778.2          & 1.56   & $\geq1.5\times10^{20}$ & $\geq1.3\times10^{19}$ & -- \\
~                    & ~             & $2^+$ 1625.9   & 847.7          & 1.96   & $\geq3.1\times10^{20}$ & $\geq1.6\times10^{19}$ & -- \\
2$\varepsilon$       & 2$\nu$        & g.s.           & --             & --     & --                     & --                     & $4.7\times10^{20}-3.9\times10^{21}$ \\
~                    & ~             & $2^+$ 778.2    & 778.2          & 2.83   & $\geq2.6\times10^{20}$ & $\geq6.5\times10^{19}$ & $4.2\times10^{28}-2.2\times10^{32}$ \\
~                    & ~             & $0^+$ 1148.1   & 778.2          & 2.64   & $\geq2.5\times10^{20}$ & $\geq4.2\times10^{19}$ & $4.2\times10^{21}-9.2\times10^{22}$ \\
~                    & ~             & $2^+$ 1497.8   & 778.2          & 1.82   & $\geq1.7\times10^{20}$ & $\geq3.0\times10^{19}$ & $1.8\times10^{29}-6.5\times10^{32}$ \\
~                    & ~             & $2^+$ 1625.9   & 848.2          & 2.29   & $\geq3.6\times10^{20}$ & $\geq3.9\times10^{19}$ & $>1.6\times10^{29}$ \\
~                    & ~             & $2^+$ 2095.8   & 778.2          & 2.56   & $\geq2.4\times10^{20}$ & $\geq4.3\times10^{19}$ & -- \\
~                    & ~             & $2^+$ 2426.1   & 778.2          & 2.28   & $\geq2.1\times10^{20}$ & $\geq3.5\times10^{19}$ & -- \\
~                    & ~             & $(0)^+$ 2622.5 & 778.2          & 2.61   & $\geq2.4\times10^{20}$ & $\geq4.6\times10^{19}$ & -- \\
2$\varepsilon$       & 0$\nu$        & $2^+$ 778.2    & 778.2          & 2.61   & $\geq2.4\times10^{20}$ & $\geq6.4\times10^{19}$ & -- \\
~                    & ~             & $0^+$ 1148.1   & 778.2          & 2.46   & $\geq2.3\times10^{20}$ & $\geq4.1\times10^{19}$ & -- \\
~                    & ~             & $2^+$ 1497.8   & 778.2          & 1.67   & $\geq1.6\times10^{20}$ & $\geq2.9\times10^{19}$ & -- \\
~                    & ~             & $2^+$ 1625.9   & 847.7          & 2.12   & $\geq3.3\times10^{20}$ & $\geq3.8\times10^{19}$ & -- \\
~                    & ~             & $2^+$ 2095.8   & 778.2          & 2.39   & $\geq2.2\times10^{20}$ & $\geq4.3\times10^{19}$ & -- \\
~                    & ~             & $2^+$ 2426.1   & 778.2          & 2.20   & $\geq2.1\times10^{20}$ & $\geq3.4\times10^{19}$ & -- \\
~                    & ~             & $(0)^+$ 2622.5 & 778.2          & 2.60   & $\geq2.4\times10^{20}$ & $\geq4.5\times10^{19}$ & -- \\
2$K$                 & 0$\nu$        & g.s.           & 2674.5         & 1.56   & $\geq1.0\times10^{21}$ & $\geq5.4\times10^{19}$ & $2.8\times10^{34}$ \cite{Ver83} \\
$KL$                 & 0$\nu$        & g.s.           & 2691.6         & 1.58   & $\geq2.3\times10^{20}$ & $\geq6.9\times10^{19}$ & -- \\
2$L$                 & 0$\nu$        & g.s.           & 2708.7         & 1.55   & $\geq2.3\times10^{20}$ & $\geq6.9\times10^{19}$ & -- \\
Resonant $KL$        & 0$\nu$        & $2^+$ 2700.2   & 1921.8         & 0.60   & $\geq2.0\times10^{20}$ & $\geq2.7\times10^{19}$ & $3.0\times10^{26}-6.0\times10^{34}$ \cite{Kri11} \\
Resonant 2$L$        & 0$\nu$        & ~~~~  2712.7   & 812.6          & 2.28   & $\geq3.6\times10^{20}$ & $\geq2.0\times10^{19}$ & $4.4\times10^{31}-2.3\times10^{32}$ \\
~                    & ~             & ~              & ~              & ~      & ~                      & ~                      & ~  \\
\multicolumn{8}{l}{$^{104}$Ru $\to$ $^{104}$Pd} \\
2$\beta^-$           & 2$\nu$        & $2^+$ 555.8    & 555.8          & 3.09   & $\geq6.6\times10^{20}$ & $\geq1.9\times10^{20}$ & $>1.8\times10^{28}$ \cite{Suh11} \\
~                    & 0$\nu$        & $2^+$ 555.8    & 555.8          & 3.06   & $\geq6.5\times10^{20}$ & $\geq1.9\times10^{20}$ & -- \\
~                    & ~             & ~              & ~              & ~      & ~                      & ~                      & ~  \\
\hline
\end{tabular}
}
\end{center}
\end{table*}

\subsection{Double electron captures in $^{96}$Ru}

Double electron captures in $^{96}$Ru lead to creation of holes in the atomic shells of $^{96}$Mo.
In the $2\nu2\varepsilon$ process, all the energy release (except the part spent on atomic
shell excitation) is taken away by two neutrinos. The energy threshold in the current measurements
(around 50 keV) does not allow to search for the deexcitation processes in the atomic shell
(which have energies less than 20 keV), thus we cannot derive limits for the g.s. to g.s. $2\nu2\varepsilon$
capture.

In neutrinoless $2\varepsilon$ capture, we suppose (as also other articles on the subject) that
the energy excess is taken away by (bremsstrahlung) $\gamma$ quanta with energy
$E_\gamma = Q_{2\beta} - E_{b1} - E_{b2} - E_{exc}$, where
$E_{bi}$ is the binding energy of $i$-th captured electron on the atomic shell, and
$E_{exc}$ is the energy of the populated (g.s. or excited) level of $^{96}$Mo.
In case of transition to an excited level, in addition to the initial $\gamma$ quantum,
other $\gamma$'s will be emitted in the nuclear deexcitation process.

We did not observe peaks with energies expected in the $2\varepsilon$ decays of $^{96}$Ru
in the experimental data. Limits on the areas of the peaks were obtained using the fitting procedure
as explained in the previous section. Fig.~5 shows an interval of the spectrum around 1921.8 keV 
and its
fit by the sum of a straight line (representing the background) and the peak with energy of 1921.8
keV expected in the $^{96}$Ru resonant decay to the 2700.2 keV level of $^{96}$Mo.

\begin{figure}[htb]
\begin{center}
\resizebox{0.54\textwidth}{!}{\includegraphics{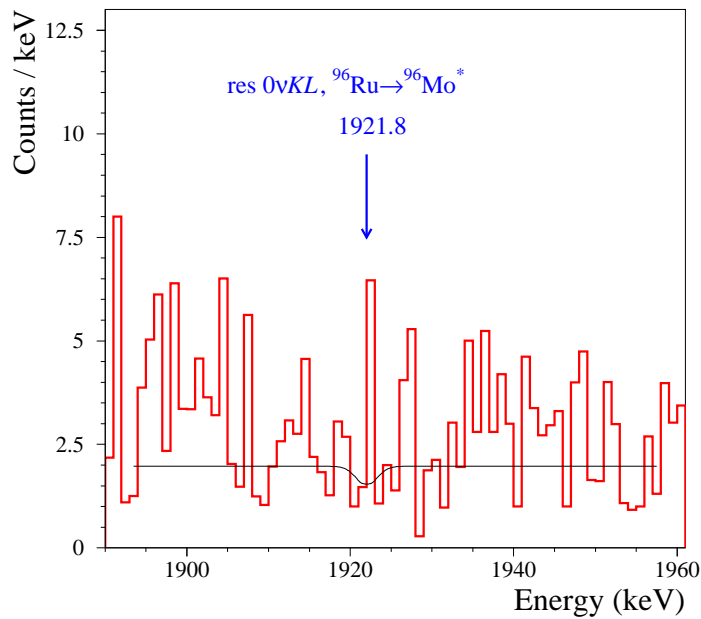}}
\end{center}
\caption{(Color on-line) Fragment of the energy spectrum accumulated with the purified Ru sample over 5479 h by
the ultra-low background HPGe $\gamma$ spectrometer. The fit is shown by solid
line. The arrow shows the energy of a peak at 1921.8 keV due to possible resonant 0$\nu KL$ capture
in $^{96}$Ru and further de-excitation of the 2$^+$, 2700.2 keV level of $^{96}$Mo.}
\end{figure}

All the obtained $T_{1/2}$ limits, together with the energies of the $\gamma$ lines
which were used to set the $T_{1/2}$ limits and corresponding detection efficiencies,
are summarized in Table~2.

\subsection{$2\beta^-$ decay $^{104}$Ru $\to$ $^{104}$Pd$^*$}

In case of $2\beta^-$ decay of $^{104}$Ru, only one excited level (2$^+$, 555.8 keV) can be
populated (see Fig.~1b). The peak at the energy 555.8 keV is absent in the experimental data (see Fig.~3).
The fit of the spectrum accumulated over 5479 h was bounded within
the $(530-600)$ keV interval. The best fit ($\chi^2$/n.d.f. = 38.2/43 = 0.89) was achieved in the
energy interval ($540-590)$ keV. The derived area of the effect, ($-17.4 \pm 18.8$) counts,
corresponds to $\lim S$ = 16.2 counts at 90\% C.L. \cite{Fel98}.
Using the number of $^{104}$Ru nuclei in the purified ruthenium sample ($N = 7.98\times10^{23}$)
and the very close detection efficiencies for the 555.8 keV $\gamma$ quanta in case of $2\nu$
and $0\nu$ mode (3.09\% and 3.06\%, respectively), the following half life limits were reached:

\begin{center}
$T_{1/2}^{2\nu 2\beta^-}$(g.s. $\rightarrow$ 2$^+$, 555.8 keV) $\geq6.6\times 10^{20}$~yr, \\
$T_{1/2}^{0\nu 2\beta^-}$(g.s. $\rightarrow$ 2$^+$, 555.8 keV) $\geq6.5\times 10^{20}$~yr.
\end{center}

\subsection{Theoretical estimates}

The theoretical estimations of Table~2 were obtained by using a 
higher-QRPA framework \cite{Suh93,Civ94} with detailed expressions given in 
\cite{Aun96,Suh98}. For the neutrinoless modes of decay the UCOM short-range 
correlations \cite{Kor07} were used. All computational details are given in a 
recent article \cite{Suh12}. Estimates coming from other sources 
\cite{Suh11,Ver83,Kri11} are indicated in the table. 
In addition, a summary of other theoretical results can be found in our previous 
work \cite{Bel09}.

\section{Discussion}

Experimental searches for $2\beta^+/\varepsilon\beta^+/2\varepsilon$ processes 
are not so popular as those for $2\beta^-$ decays. There are three reasons for 
such a situation:

(1) These nuclei mostly have low natural abundance, usually less than 1\%, with few
exceptions \cite{2bTab} (and $^{96}$Ru with $\delta=5.54\%$ is among them);

(2) Energy available for positrons is related with the energy release $Q_{2\beta}$ as
$Q_{2\beta}-4m_ec^2$ (for $2\beta^+$ decay) or $Q_{2\beta}-2m_ec^2-E_b$ (for $\varepsilon\beta^+$),
where $m_ec^2$ is the electron rest mass, and $E_b$ is the binding energy of the captured electron 
on the atomic shell. This leads to smaller phase space factors in comparison with $2\beta^-$ decay, 
and thus in lower probabilities for $2\beta^+/\varepsilon\beta^+$ processes;

(3) In searches for X rays emitted in deexcitation of atomic shells in case of 
$\varepsilon\beta^+/2\varepsilon$ decays, detectors with low energy threshold (and good energy
resolution) are needed; in addition, it is difficult to ensure high efficiency for 
detection of low energy X rays when external $2\beta$ sources are investigated.

In result, while in searches for neutrinoless $2\beta^-$ decay the sensitivity of
$T_{1/2}>10^{25}$ yr was achieved (for $^{76}$Ge \cite{Ge76} and $^{136}$Xe \cite{Xe136}), 
the best $T_{1/2}$ limits achieved in $2\beta^+/\varepsilon\beta^+/2\varepsilon$ experiments 
are much more modest.
Sensitivity $T_{1/2}>10^{20}$ yr was reached in direct experiments for
$^{54}$Fe \cite{Bik98},
$^{58}$Ni \cite{Vas93},
$^{64}$Zn \cite{Bel11a},
$^{92}$Mo \cite{Lee11},
and limits $T_{1/2}>10^{21}$ yr were obtained for
$^{40}$Ca \cite{Bel99},
$^{78}$Kr \cite{Sae94},
$^{106}$Cd \cite{Bel12},
$^{112}$Sn \cite{Bar11},
$^{120}$Te \cite{And11},
$^{132}$Ba \cite{Mes01}.
Geochemical experiments currently give an indication on $2\nu2\varepsilon$ capture 
in $^{130}$Ba with
$T_{1/2}=(2.2\pm0.5)\times10^{21}$ yr \cite{Mes01} and 
$T_{1/2}=(6.0\pm1.1)\times10^{20}$ yr \cite{Puj09}
(also limit $>4.0\times10^{21}$ yr is known \cite{Bar96}).
In addition, an observation of $2\nu2K$ capture in $^{78}$Kr was recently claimed; 
the obtained half life is
$T_{1/2}=1.4^{+2.2}_{-0.7}\times10^{22}$ yr (however, also cautious limit is given as
$T_{1/2}>7.0\times10^{21}$ yr at 90\% C.L.) \cite{Gav11}.

As for resonance $0\nu2\varepsilon$ capture, intensive high-precision measurements of
$Q_{2\beta}$ values during last few years (see reviews \cite{Eli12} and refs. therein)
excluded many nuclei from the list of perspective candidates in searches for this exotic process,
leaving in the list only $^{152}$Gd and $^{156}$Dy. 
While for $^{152}$Gd experimental investigations were not performed to-date, 
for $^{156}$Dy first experimental limits \cite{Bel11b} were set on the level of only 
$T_{1/2}>10^{16}$ yr (this is related with $^{156}$Dy low natural abundance 
$\delta=0.056\%$ \cite{Ber11}).

Comparing the above described $T_{1/2}$ limits for different isotopes with the 
values obtained in the present measurements, one could conclude that 
the latter are on the level of the best results achieved to-date in other experiments.

\section{Conclusions}

A low background experiment to search for 2$\beta$ processes in $^{96}$Ru and $^{104}$Ru isotopes
was carried out over more than 7.6 thousands hours in the underground Gran Sasso National Laboratories
of the I.N.F.N. measuring ruthenium samples with ultra-low background HPGe 
detectors. The total exposure of the experiment is
0.56 kg $\times$ yr. Purification of the ruthenium using the electron beam melting method allowed
to reduce the potassium contamination by more than 20 times; activities of $^{226}$Ra and $^{106}$Ru were 
decreased as well.

The new improved half life limits on double beta processes in $^{96}$Ru have been set at
the level of $10^{20} - 10^{21}$~yr. Moreover, the 2$\beta^-$ transition of $^{104}$Ru to the excited 2$^+$ level
of $^{104}$Pd has been investigated with the same sensitivity.
All results give higher values than those recently published  \cite{Bel09,Bel10,And12}.
However, the limits are still far from the theoretical predictions,
with the exception of the $2\nu\varepsilon\beta^+$ channel for $^{96}$Ru (g.s. to g.s. transition)
and $2\nu2\varepsilon$ decays with population of the $0^+$ levels (g.s. and the first
$0^+$ level with $E_{exc}=1148.1$ keV),
for which half lives $T_{1/2} \simeq 10^{21} - 10^{22}$ yr have been estimated.

\section{Acknowledgments}

The group from the Institute for Nuclear Research (Kyiv, Ukraine) was supported in part by
the Space Research Program of the National Academy of Sciences of Ukraine.
We thank anonymous referee for useful suggestions.

\end{document}